\documentclass[article]{elsarticle}
\usepackage{hyperref,amssymb}

\journal{Nuclear Physics A}

\newcommand{\bs}{\begin{subequations}}
\newcommand{\es}{\end{subequations}}

\newcommand{\be}{\begin{equation}}
\newcommand{\ee}{\end{equation}}

\newcommand{\el}{\left}
\newcommand{\er}{\right}

\newcommand{\dis}{\displaystyle}

\begin{document}

\begin{frontmatter}

\title{Analysis of the Pion-Nucleus Scattering within the Folding and the Kisslinger Type Potentials}

\author[1]{V.K.Lukyanov\corref{mycorrespondingauthor}}
\cortext[mycorrespondingauthor]{Corresponding author.}
\ead{vlukyanov@jinr.ru}

\author[1,2]{E.V.Zemlyanaya}
\author[1,2]{K.V.Lukyanov}
\author[3]{I.Abdul-Magead}

\address[1]{Joint Institute for Nuclear Research, Dubna 141980, Russia}
\address[2]{Dubna State University, Dubna 141980, Russia}
\address[3]{Cairo University, Giza, Cairo, Egypt}

\begin{abstract}
Elastic cross sections are calculated and compared with the data on $\pi^{\pm}$ scattering on $^{28}$Si, $^{40}$Ca, $^{58}$Ni, $^{208}$Pb nuclei in the energy range from 130 to 290 MeV. To this end, both the folding optical potential (OP) and the local modified Kisslinger-type OP were calculated, and then the $\pi A$ cross sections were obtained  by solving the Klein-Gordon equation to account for the relativization and distortion wave effects. In the folding OPs, the parameters of the elementary $\pi N$ amplitude were fitted when describing the $\pi A$ scattering data, and thus the essential in-medium effect on the parameters of the $\pi N$ amplitude  was established since the pion is scattered not on a free but on a bound nuclear nucleon. Fairly good agreement with experimental data was obtained for both models of optical potentials and their forms turn out to be in coincidence in the region of their surfaces.
\end{abstract}


\end{frontmatter}


\section{Introduction}

In the present work, basing on two models of microscopic  optical potentials we consider a wide set of experimental data on the pion-nucleus ($\pi A$) elastic scattering at energies from about 100 to 300 MeV where the pion-nucleon  $p$-wave $\Delta_{33}$-resonance reveals itself. One of the models is the folding $\pi A$ potential where the nuclear density distribution function is integrated with the $\pi N$ amplitude in the form depending on its observed characteristics of scattering. The other model is the Kisslinger $\pi A$ potential in its equivalent local form obtained  by using the Krell-Ericson transformation.

It should be mentioned that there exists a considerable amount of theoretical investigations
of the ($\pi A$) scattering  where nuclear density distribution functions and the $\pi N$ scattering amplitude are used in calculations of the pion-nucleus cross sections.
So in the early papers \cite{Faldt68} and \cite{Bjornenak70}, the data of elastic scattering on the deuteron and the light nuclei $^{12}$C,$^{51}$V,$^{90}$Zr  were considered using the Glauber multiple-scattering theory (see, e.g. Refs. \cite{Glauber59} and \cite{Czyz69}), and  thus, the conclusions were made that the Fermi motion of nucleons in the target nuclei has a large influence on the results.  Later on, many approaches were developed for constructing first-order optical potentials for calculations of the $\pi A$ cross sections and comparing them with the respective  experimental data. Thus, e.g., in \cite{Charlton71}, a separable form for the $\pi$-nucleon interaction was suggested, and then the respective calculations were compared with those adjusted in the  general multiple-scattering approaches. Also, in Ref.\cite{Landau73}, basing on the multiple scattering theory, three optical pion-nucleus potentials were  studied in momentum space. The first one is the  non-local Kisslinger optical potential \cite{Kiss55} of the form $a\rho(r)+b\nabla(\rho(r)\nabla)$, the second one  is its simplified local form $a\rho(r)+b\nabla^2\rho(r)$, and the third one ("Laplacian form")  includes additional terms proportional to $(\nabla^2)^n\rho(r)$ (n=2,4). The detailed applications of the local form to the pion elastic and inelastic scattering on the nuclei $^{12}$C,$^{58}$Ni, $^{176}$Yb at energies of the 33-resonance can be see in Ref.\cite{Miller74}. In Ref.\cite{Liu81} when discussing higher order potentials, the theory was based upon the use of the analyticity and unitarity properties of the $\pi A$ scattering amplitude, which deviates from the description of high-order terms of the standard multi-scattering theory. This helps to construct a first-order optical potential that enables one to fit the corresponding calculations with some selected elastic scattering data.  The angular distributions for elastic scattering of pions by $^{40,42,44,48}$Ca at three energies 116,180 and 292 MeV were analyzed  in Ref. \cite{Khallaf2002}. It was established that the skin neutrons in $^{44,48}$Ca contribute to $\pi$ meson elastic scattering in the energy range of the 33-resonance  more than protons.

The other approach is the calculations of the pion-nucleus elastic scattering cross sections based on the Woods-Saxon optical potential. In particular, it was established in Ref.\cite{Satchler92} that the nuclear surface region plays a decisive role in fitting calculations to experimental data. Also, when analyzing  the obtained there best fitted parameters of the optical potential for $\pi^{\pm}$ scattering on $^{208}$Pb at energies of 116, 162, 180 and 291 MeV (see Table 1 of \cite{Satchler92}), one can find that the corresponding $\sigma_R$ cross sections reveal a maximum at about 180 MeV in the 33-resonance energy region.

Also, one should mention a microscopic consideration of the problem in  Ref.\cite{Kamalov87} where the introduced ${\pi A}$ optical potential (OP), being proportional to the first-order nuclear density $\rho(r)$, was supplemented by the squared density term $\rho^2(r)$  that defines the two-body-like character of pion  absorbtion around the nuclear surface.  This term was taken to be proportional to the energy-dependent complex parameters $B_0(E)$ and $C_0(E)$, which contribute to the depth of the folding OP. As one can see from Table 1 in                                     f Ref.\cite{Kamalov87}, the imaginary parts of these parameters, being proportional to the imaginary parts of OPs, have a maximum at a pion energy at about 180 MeV, just near the maximum of the total  $\pi A$ cross section.

When constructing the microscopic $\pi A$ optical potential at energies near the $\Delta_{33}$-resonance, its physical nature was firstly considered to be connected with the ordinary non-local Kisslinger potential \cite{Kiss55} of the form $a(r)\,+\,\nabla b(r)\nabla$  with the specific p-wave term. Over the years, this non-local form was transformed in a local potential with the help of the Krell-Ericson   transformation \cite{KrellEr69} of an ordinary wave function that obeys the Klein-Gordon equation. As the final result, one obtains the Shroedinger equation with the transformed local $\pi A$ potential of a rather complicated form in Ref. \cite{JohnSatch96}.

In our study we will apply this form of  OP in calculations of the $\pi^{\pm}$ elastic scattering cross sections on nuclei in the 33-resonance energy region. This OP depends on 12 parameters for calculations of pion scattering on a number of different target-nuclei but at the same collision energy.

At the same time, we are going to apply the folding $\pi A$ optical potential from Ref. \cite{LukZemLuk06}, which depends on three parameters of the elementary pion-nucleon scattering amplitude. One of them is $\sigma$, the total pion-nucleon  cross section, the other is $\alpha$, the ratio of the real to imaginary parts of the $\pi N$ amplitude at forward angles, and the third parameter $\beta$ defines the slope of the pion-nucleon amplitude at small scattering angles. When analyzing the $\pi A$ scattering data these $\pi N$  parameters  reveal their specific dependence on the pion energies near the 33-resonance.

The first aim of our study is to calculate, in the framework of both models,  the pion-nucleus elastic scattering cross sections at energies in the 33-resonance energy region and then to compare them with the respective set of experimental data. The second aim is a study, within the folding model, the ``in-medium''  effect when the pion is scattered on a bound nuclear  nucleon but not on a free one. Making use of the fact that in the folding model the fitted parameters are the $\pi N$ total cross section $\sigma$ and the ratio $\alpha$ of the real to imaginary parts of the $\pi N$ scattering amplitude at zero angles, we  compare these ``in-medium'' characteristics with the respective ``free'' ones.   Below, calculations are presented for pion elastic scattering on nuclei $^{28}$Si, $^{40}$Ca, $^{58}$Ni and $^{208}$Pb at energies of 130, 162, 180, 226, and 291 MeV in the region of the 33-resonance energy.

\section{Basic equations}
$The \,\, \pi A \,\, microscopic  \,\, OP$ is considered starting from its folding form in the momentum representation
\begin{equation}\label{c.1}
U(r)\,=\, {1\over (2\pi)^3}\,\int {e^{\dis -i{\bf q}{\bf r}}} \,\rho(q)\,v_{{\pi}N}(q)\,d^3q\,=\,
{1\over 2\pi^2}\,\int j_0(qr)\,\rho(q)\,v_{{\pi}N}(q)\,q^2\,dq.
\end{equation}
Here the form factors $\rho(q)$ and $v_{{\pi}N}(q)$ correspond to the target nucleus density distribution and the pion-nucleon potential. The latter
\begin{equation}\label{c.2}
v_{{\pi}N}(q)\,=\,\int e^{\dis i{\bf q}{\bf r}} \,v_{{\pi}N}(r)\,d^3r
\end{equation}
can be expressed through the ${\pi}N$  scattering amplitude in the Born approximation
\begin{equation}\label{c.3}
F^B_{{\pi}N}(q)\,=\,-\,{m_{\pi}\over 2\pi\hbar^2}\,\int e^{\dis i{\bf q}{\bf r}} \,v_{{\pi}N}(r)\,d^3r\,=\,-\,{m_\pi\over 2\pi\hbar^2}\,v_{{\pi}N}(q),
\end{equation}
where $m_{\pi}$ is the mass of an incident ${\pi}$-meson. On the other hand, the ${\pi}N$ amplitude can be represented in the form that corresponds to the optical theorem
\begin{equation}\label{c.4}
F_{{\pi}N}(q)\,={k\over 4\pi}(i\,+\,\alpha)\,\sigma \,f_{\pi N},\qquad    f_{\pi N}(q)\,=\, e^{-\beta \,q^2/2}.
\end{equation}
Here $\sigma$ is the $\pi N$ total cross section, $\alpha$ is the ratio of real to imaginary parts of the amplitude of scattering at zero angles and $\beta$ is the slope parameter. When equating two $\pi N$ amplitudes (3) and (4), one obtains
\begin{equation}\label{c.5}
v_{\pi N}(q)\,=\,-{\hbar v\over 2}\, (i\,+\,\alpha)\,\sigma \, f_{\pi N}(q),
\end{equation}
and then substituting (5) into (1) one gets  the pion-nucleus optical potential Ref.\cite{LukZemLuk06}
\begin{equation}\label{c.6}
U(r)\,=\,-(\hbar c)\cdot\beta_c {\sigma\over (2\pi)^2}\,
(\alpha\,+\,i) \,\int \,j_0(qr)\,\rho(q)\,f_{\pi N}(q)\,q^2dq,
\end{equation}
where one has the $\pi$ meson velocity in the $\pi A$ center-of-mass system $\beta_c =v/c=\,{\sqrt{T_{lab}(T_{lab}+2m_{\pi}})}/[T_{lab}+m_{\pi}\cdot {M_A+m_{\pi}\over M_A}]$ with the incident $\pi$ meson kinetic energy $T_{lab}$, and the target nucleus mass $M_A$.

When calculating the folding OP (6), we apply the nuclear density distribution function $\rho(r)$ of  point-like nucleons in the form of a symmetrized Fermi-function (SF-function)
\begin{equation}\label{c.7}
\rho_{SF}(r)=\rho_0\frac{\sinh\left( R/a \right)}
        {\cosh\left( R/a \right)+\cosh\left( r/a \right)},\quad \rho_0= {\frac {A} {1.25\pi R^3}} {\left[1+(\frac{\pi a}{R})^2 \right]^{-1}},
\end{equation}
whose form factor has the explicit form
\begin{equation}\label{c.8}
\rho_{SF}(q)\, =\, - \rho_0\, {4\pi^2aR \over q}\,
{\cos qR\over \sinh(\pi aq)}
\el[1 -\el({\pi a\over R}\er)\coth(\pi aq)\tan qR\er].
\end{equation}
One should remind that the parameters $R$ and $a$ in (8) correspond to the nuclear density distribution $\rho_{SF}(r)$ of  $\it point\,-\,like\, nucleons$, and they are known from the respective tables. Thus, when calculating the optical potentials (6) and then the corresponding cross sections, we compare the latter with experimental data by fitting only three parameters $\sigma$, $\alpha$, and $\beta$. These parameters  have an obvious physical meaning and characterize  the pion scattering on the $bound \,nuclear \, nucleons$, and thus one can say of the so-called  $"in-medium"\, effect$ on the $\pi N$ amplitude of scattering.

$The \, local \, model \,of \, Kisslinger-type \,\,OP$  was realized in Ref.\cite{JohnSatch96}  in the form
\begin{eqnarray}
 \qquad\qquad\qquad   V(r)={(\hbar c)^2\over 2E}{1\over {1-\alpha}} \Biggl\{u - k^2\alpha - \Biggl[{1\over 2}\nabla^2\alpha +
    {\bigl({1\over 2}\nabla\alpha\bigr)^2\over{1-\alpha }}\Biggr]\Biggr\},
\end{eqnarray}
where the total energy $E=T_{cm}+ m_\pi$ consists of the c.m. kinetic energy and the pion mass. Then, the first term $q(r)$ arises from the s-wave and partly the p-wave $\pi N$ interactions, while other smaller terms are from the p-wave contributions only. Here they are given in the form as in \cite{JohnSatch96} at  pion energies larger than a hundred MeV
\begin{eqnarray}
    u(r)=-4\pi \left[{\bf b_0}\rho(r)\mp {\bf b_1}\Delta\rho(r)\right]+\delta q, ~~~~~~~~~~~~~~~~~~~~~~~~~~~~~~~~~~~~~~
\end{eqnarray}
\begin{eqnarray}
    \alpha(r)=4\pi\left[{\bf c_0}\rho(r)\mp {\bf c_1}\Delta\rho(r)\right]/p_1+4\pi\left[{\bf C_0}\rho^2(r)\mp {\bf C_1}\rho(r)\Delta\rho(r)\right]/p_2,~~~~~
\end{eqnarray}
\begin{eqnarray}
    \delta q(r)=-2\pi\epsilon\nabla^2\biggl\{\left[{\bf c_0}\rho(r)\mp {\bf c_1}\Delta\rho(r)\right]/p_1+
    {1\over2}\left[{\bf C_0}\rho^2(r)\mp {\bf C_1}\rho(r)\Delta\rho(r)\right]/p_2]\biggr\}.
\end{eqnarray}
Here  $\bf {b_0,b_1, \, c_0,c_1, \,C_0,C_1}$ are the complex fitted parameters (the whole number 12). Their real and imaginary parts stand for the real and imaginary parts of an optical potential, respectively. They depend on the collision energy and mass of the target nucleus. Then, $p_1=1+\epsilon$, $p_2=1+(1/2)\epsilon$ with $\epsilon=E/M $, where $E$ is the total energy and $M$ is the nucleon mass.

The differential cross sections are calculated as in \cite{LukZemLuk10} by  solving the Klein-Gordon equation in its form at the conditions $E\gg U$, where $E = \sqrt{k^2+m_\pi^2}$ is the total pion energy. Then
\begin{equation}
        \left(\Delta + k^2 \right) \psi({\vec{r}}) = 2 \bar\mu U (r) \psi({\vec{r}}), \quad U(r) = V(r) + V_C(r).
\end{equation}
Here $k$ is the relativistic momentum of a pion in the center-of-mass (c.m.) system,
\begin{equation}
        k= \frac{M_A k^{lab}}{\sqrt{(M_A+m_\pi)^2+2M_A T_{lab}}},\quad
        k^{lab}= \sqrt{ T_{lab} \left( T_{lab} + 2m_{\pi}\right)},
\end{equation}
and $\bar\mu=\frac{E M_A}{E+M_A}$  is the relativistic reduced mass of the $\pi + A$ system,
which in fact is the total energy of the $m_\pi + M_A$ center-of-mass (c.m.) system.

In the case of the folding potential (6), when comparing calculated pion-nucleus cross sections with experimental data, the fitted parameters $\sigma$, $\alpha$, and $\beta$ have a meaning of the so-called  "in-medium" parameters for pion scattering on bound nucleons. The fitting procedure was realized by minimization of
the $\chi^2$-function
\begin{equation}
  \chi^2 =   f \left( \sigma, \alpha, \beta \right)
  =\sum_{i}{ \frac{\left[ y_i - \hat{y_i}(\sigma,\alpha, \beta)\right]              ^2}{s_{i}^2}},
\end{equation}
where  $y_i= {\displaystyle\frac{d\sigma_i}{d\Omega}}$ and $\hat{y_i}= {\dis\frac{d\sigma_i}{d\Omega}}(\sigma,\alpha, \beta)$ are, respectively, experimental and theoretical differential cross sections and $s_i$ are experimental errors. In this work, when performing minimization of the fitting parameters $\sigma,\alpha, \beta$, we use the ordinary cross sections  ${\displaystyle\frac{d\sigma}{d\Omega}}$ while in the preceding works \cite{LukZemLukZZ1214} minimization was made of the functions $\log\Bigl({\displaystyle\frac{d\sigma}{d\Omega}}\Bigr)$. Also, in the preceding study  the data were fitted at all available angles of scattering, different at different energies of $\pi$ mesons. However, in the present study to exclude this  kind of uncertainty in the $\sigma,\alpha, \beta$ parameters, we consider the experimental information in the same region of angles of scattering lower than 80 degrees.

\section{Results of calculations}

\begin{figure}[!ht]
\centering
\includegraphics[width = 0.9\linewidth]{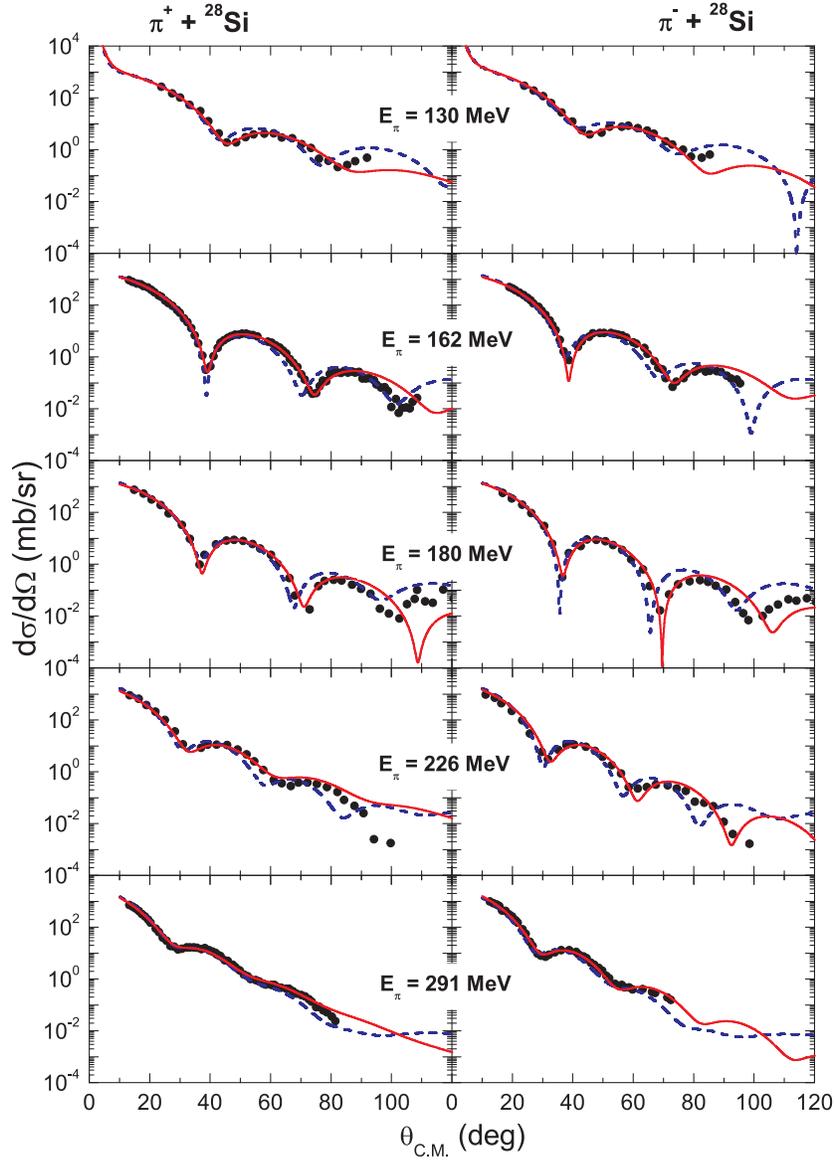}
\caption{Comparisons of the calculated ${\pi^{\pm}}+{^{28}}$Si elastic scattering cross sections at pion energies of 130,162,180,216 and 230 MeV  with the experimental data from \cite{{Preedom79},{Olmer80},{Geesaman81}} . The solid (red) curves correspond to the folding OP, and the dashed (blue) curves are for the local Kisslinger-type OP.  The best fit "in-medium" parameters $\sigma, \alpha, \beta$ of the $\pi N$ amplitude of the folding OPs are from Table 1, and parameters of the Kisslinger-type OPs are from Table 2.}
\end{figure}
%
%
\begin{figure}[ht]
\centering
\includegraphics[width = 0.9\linewidth]{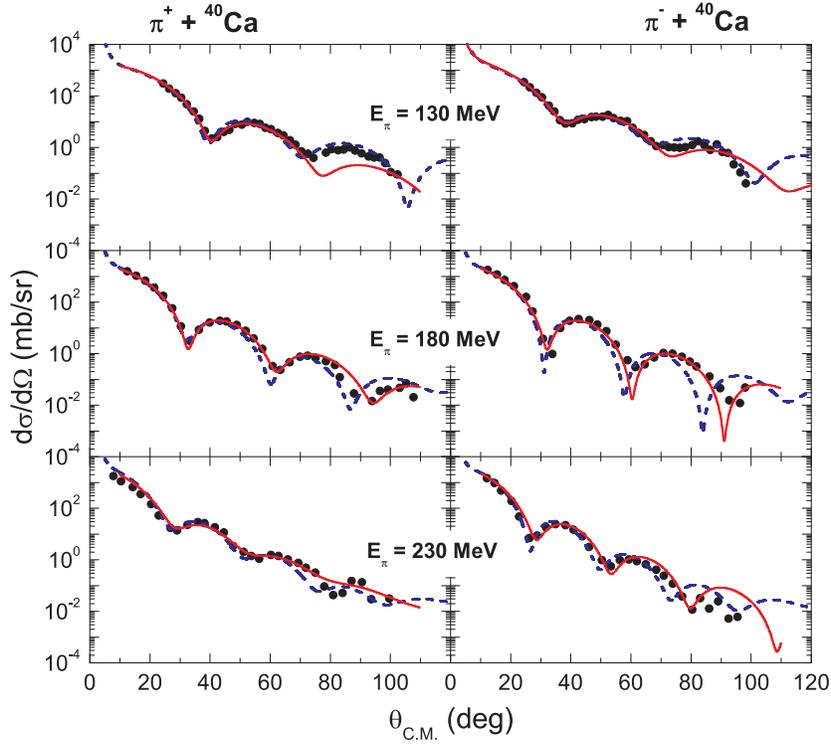}
\caption{The same as in Fig.1 but for the elastic scattering of ${\pi^{\pm}}+{^{40}}$Ca at pion energies of 130,180 and 230 MeV. Experimental data are from \cite{Gretillat81}.}
\end{figure}
%
%
\begin{figure}[!ht]
\centering
\includegraphics[width = 0.9\linewidth]{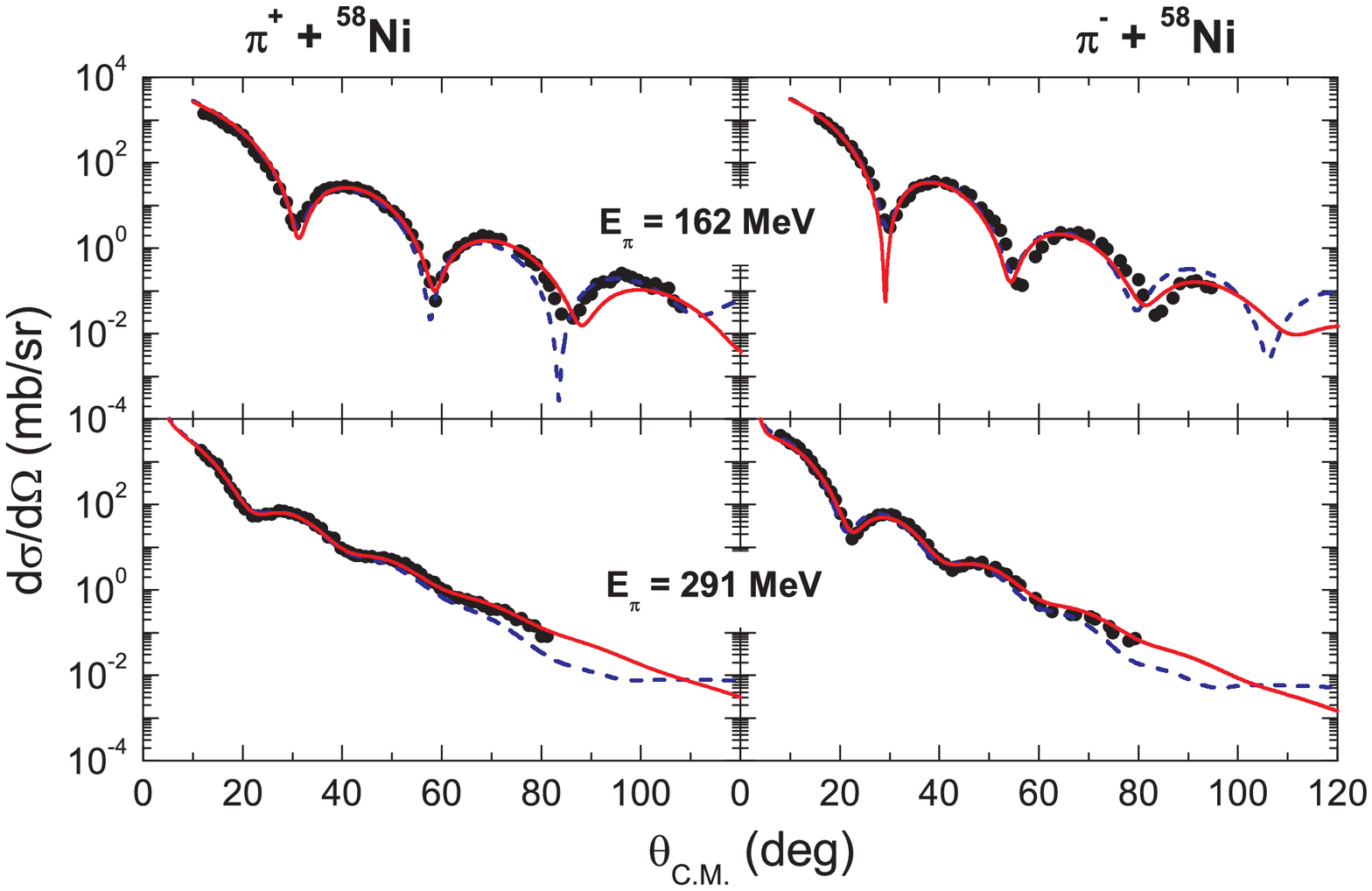}
\caption{The same as in Fig.1 but for the elastic scattering of ${\pi^{\pm}}+{^{58}}$Ni at pion energies of 162 and 291 MeV. Experimental data are from \cite{{Olmer80},{Geesaman81}}.}
\end{figure}
%

%
\begin{figure}[ht]
\centering
\includegraphics[width = 0.9\linewidth]{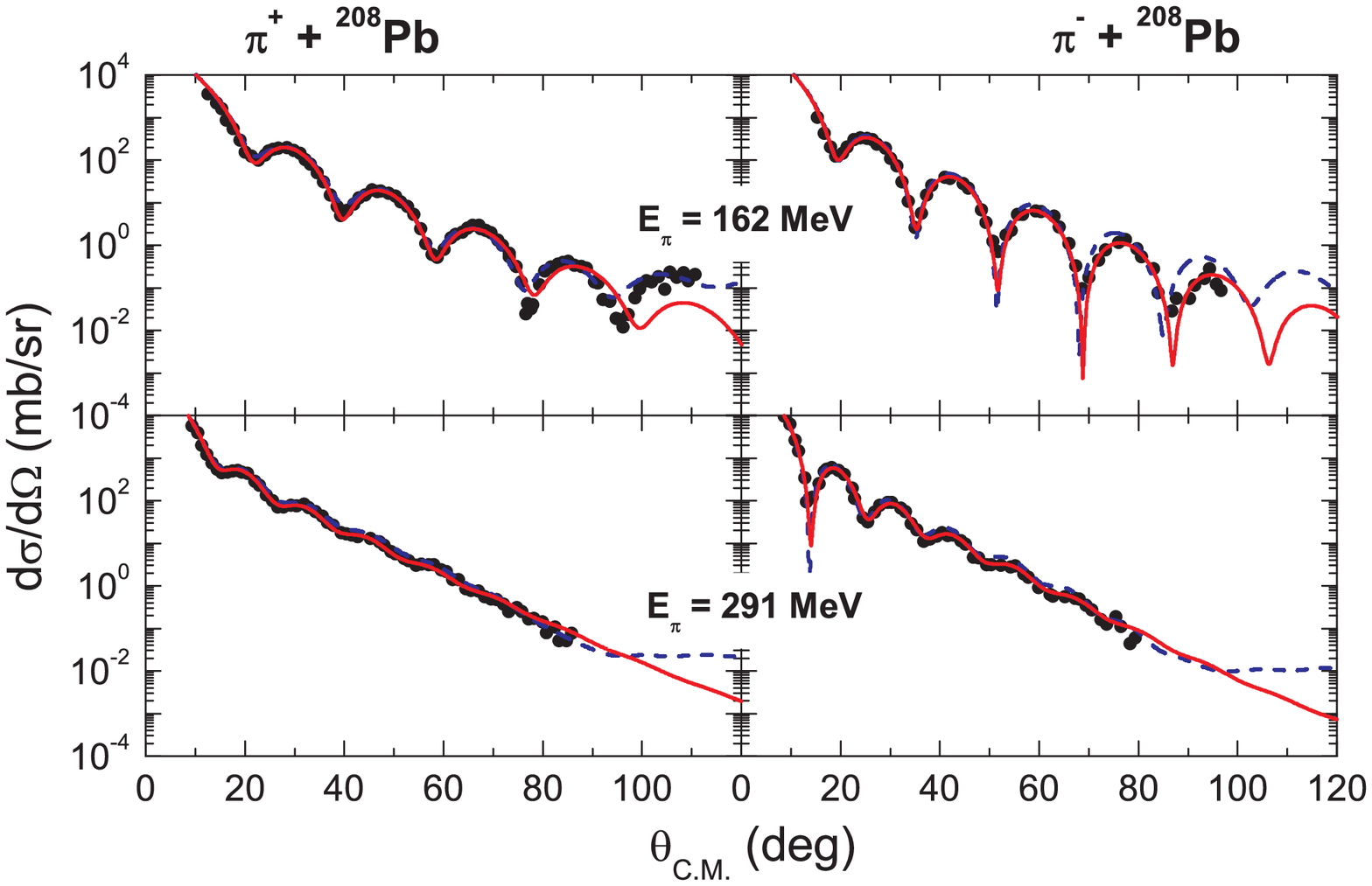}
\caption{The same as in Fig.1 but for the elastic scattering of ${\pi^{\pm}}+{^{208}}$Pb at pion energies of 162 and 291 MeV. Experimental data are from \cite{{Olmer80},{Geesaman81}}.}
\end{figure}

%
\begin{table}[!ht]
\centering
\caption{The best-fit parameters $\sigma$, $\alpha$, $\beta$ and respective  $\chi^2/k$ quantities where $k$ is the number of experimental points at $\Theta_{C.M.}$ between 0  and  80 degrees.}

\begin{tabular}{|l|c|c|r|c|c|}
 \hline
reaction&	$T^{\rm lab}$, MeV  &  $\sigma$, fm$^{2}$ &  $\alpha$ & $\beta$, fm$^{2}$ & $\chi^2/k$ \\
\hline	
$\pi^-$+$^{28}$Si& 130      &7.00  &0.84  &0.93  & 0.122 \\
$\pi^+$+$^{28}$Si&          &5.87  &1.00  &1.42  & 0.872 \\
$\pi^-$+$^{40}$Ca&          &7.15  &0.84  &1.02  & 0.603 \\
$\pi^+$+$^{40}$Ca&          &5.65  &0.98  &1.55  & 1.982 \\
\hline
$\pi^-$+$^{28}$Si& 162     &9.05   &0.49  & 0.50  & 0.398 \\
$\pi^+$+$^{28}$Si&         &8.15   &0.68 & 0.81 & 0.776 \\
$\pi^-$+$^{58}$Ni&         &10.53  &0.09  & 0.99  & 3.685 \\ 
$\pi^+$+$^{58}$Ni&         &8.03   &0.49  & 0.87  & 1.111 \\ 
$\pi^-$+$^{208}$Pb&         &10.00  &0.30  & 1.01  & 0.644 \\ 
$\pi^+$+$^{208}$Pb&         &6.22   &0.61  & 1.28  & 0.719 \\ 
\hline
$\pi^-$+$^{28}$Si& 180     &8.97  &0.37  & 0.44  & 1.262 \\
$\pi^+$+$^{28}$Si&         &8.35  &0.61  & 0.52  & 1.024 \\
$\pi^-$+$^{40}$Ca&         &8.95  &0.40  & 0.45  & 0.891 \\
$\pi^+$+$^{40}$Ca&         &6.85   &0.94 & 0.67 & 1.588 \\
\hline
$\pi^-$+$^{28}$Si& 226     &7.35   &0.54   & 0.33  & 4.160 \\
$\pi^+$+$^{28}$Si&         &8.75   &-0.17 & 0.36 & 4.997 \\
\hline
$\pi^-$+$^{40}$Ca& 230      &6.75  &0.64  & 0.29 & 2.123 \\
$\pi^+$+$^{40}$Ca&          &8.05   &-0.25 & 0.55 & 2.743 \\
\hline
$\pi^-$+$^{28}$Si& 291      &4.86  &-0.80  & 0.39  & 1.169 \\
$\pi^+$+$^{28}$Si&          &5.24  &-0.76  & 0.45  & 0.874 \\
$\pi^-$+$^{58}$Ni&         &4.55   &-0.86  & 0.32  & 0.863 \\ 
$\pi^+$+$^{58}$Ni&         &5.47   &-0.65  & 0.37  & 0.558 \\ 
$\pi^-$+$^{208}$Pb&         &4.97  &-0.93  & 0.64  & 0.592 \\ 
$\pi^+$+$^{208}$Pb&         &6.04  &-0.43  & 0.64  & 0.574 \\ 
\hline
\end{tabular} %
\end{table}
%

%
%
\begin{figure}[ht]
\centering
\includegraphics[width = 0.95\linewidth]{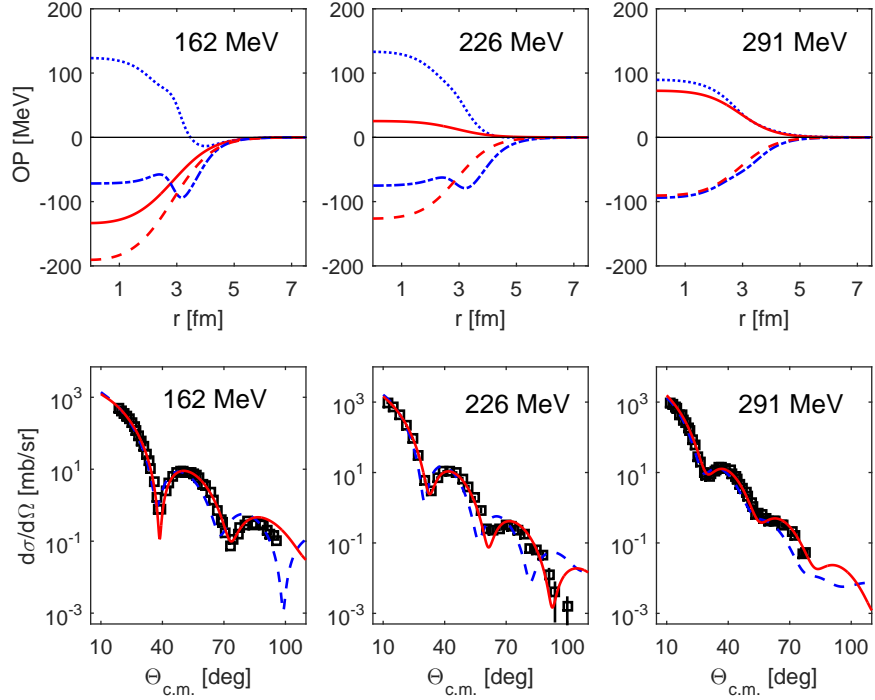}
\caption{In the first row, the folding OPs (red solid curves - real and long-dashed - imaginary parts) are compared with the Kisslinger potentials (blue dots - real and dot-dashed - imaginary parts). The second row - the  corresponding calculated cross sections (solid - by folding OPs and dashed - by Kisslinger OPs) for the  ${\pi^+}+{^{28}}$Si  scattering at T$_{lab}$ = 162, 226 and 291 Mev, as they are presented in Fig.1.}
\end{figure}
%
\begin{figure}[ht]
\centering
\includegraphics[width=.95\linewidth]{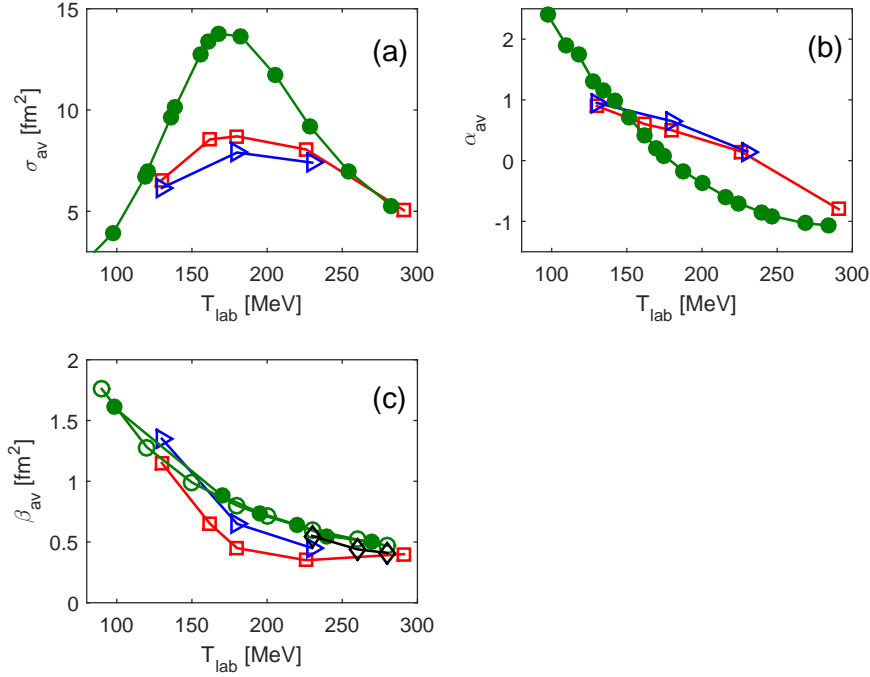}
\caption{The energy dependences of the fitted  parameters $\sigma_{av}, \alpha_{av}, \beta_{av}$  of the ''in-medium'' $\pi N$ amplitude (4), obtained from the data on $\pi +{^{28}}$Si ($\square$ red) and on $\pi + {^{40}}$Ca ($\triangleright$ blue), are compared with those from  pion scattering on ''free'' nucleons:  (a) experimental data ($\bullet$ green)  are from \cite{Carter71} , (b) -- from \cite{Arndt85} ($\bullet$ green) , (c) -- from   \cite{Locher71} ($\bullet$ and $\circ$ green) and from \cite{Bano}($\diamondsuit$ black).}
\end{figure}
When calculating differential cross sections by using the folding OP (6), we intend, first, to get their agrement  with a large set of experimental data by fitting the $\sigma$, $\alpha$, $\beta$ parameters, which define the $\pi N$  scattering amplitude on the $bound \,nuclear\, nucleons$. Later on, we will compare the energy dependence of these parameters in the 33-resonance region with their known energy dependence but for pion scattering on $free\, nucleons$.

Otherwise, when using the Kisslinger-type microscopic optical potential (9), which takes properly into account its specific at the 33-resonance energies, we fit its 12  parameters to get regular trends of their behavior with increasing collision energy. In both approaches we apply  the symmetrized fermi (SF) functions (7) for  nuclear density distributions in the case of the Kisslinger-type OP and also for distributions of  "point-like" nuclear nucleons when using the folding OP.

Below in Figs.1-4 we present the results of calculations of the pion elastic scattering
cross sections at different energies on $^{28}$Si, $^{40}$Ca,  $^{58}$Ni and $^{208}$Pb nuclei. Figure 1 demonstrates for both approaches reasonable agreement of the calculated and experimental differential cross  sections of the  $\pi^\pm$-meson elastic scattering on  $^{28}$Si at a large set of energies 130, 162, 180, 226, 230 and 291 MeV which cowers the whole region of the 3-3 resonance energy. The experimental data are from Refs.\cite{{Preedom79},{Olmer80},{Geesaman81}}. The solid red curves are the calculated cross sections for the folding OP (6) where one applies  the SF-density distribution function of point-like nucleons to the nucleus $^{28}$Si with the radius and diffuseness parameters $R_c=3.134$ fm and $a_c=0.477$ fm (see Refs.\cite{{LukZemSlow04},{LukZemLukMag17}}). The dashed $blue$ curves correspond to the cross sections calculated by using the Kisslinger type microscopic  OP, where the nuclear density distribution has the same form (7) but with the parameters  $R=3.14$ fm and $a=0.537$ fm from Ref.\cite{DeVries87} .

Figure 2 shows for the target-nucleus $^{40}$Ca the cross-section calculations based on the folding and Kisslinger type OPs and their comparison with the data of scattering at 130, 180 and 230 MeV (see Ref.\cite{Gretillat81}). For this nucleus, when calculating folding OPs we use the  point-like nucleon fermi density parameters $R_c=3.593$ fm, $a_c=0.493$ fm  (solid red curves) \cite{LukZemSlow04}, and for the Kisslinger OPs (dashed blue curves), the parameters $R=3.53$ fm and $a=0.562$ fm for distribution of realistic nuclear nucleons \cite{LukZemLukMag17}  .

Then, Fig.3 demonstrates a similar comparison of the cross sections but calculated for the nucleus $^{58}$Ni and  pion energies 162 and 291 MeV (see Refs.\cite{{Olmer80},{Geesaman81}}). Here the nuclear fermi density parameters for distributions of point-like nucleons (solid curves) are $R_c=4.08$ fm, $a_c=0.515$ fm \cite{DaoNP2000}. The parameters for realistic density distributions of $^{58}Ni$  (dashed curves) are $R=4.024$ fm and $a=0.582$ fm.

Figure 4 shows the cross section calculations by using the folding and Kisslinger type OPs at the same energies 162 and 291 MeV but for the $^{208}$Pb nucleus (the data are from  Refs.\cite{{Olmer80},{Geesaman81}}). In calculations (see also Ref.\cite{LukZemLukMag20}) there
were used  point-nucleon density parameters  $R_c=6.654$ fm, $a_c=0.475$ fm  (solid curves) \cite{PatPet2003}, while for distributions of  realistic nucleons the parameters                                                                                                                                                                                                                                                                                                                                                                                                                                                                                                                                                                                                                                                                                                                                                                                                                                                                                                                                                                                                                                                                                                                                $R=6.67$ fm, $a=0.545$ fm (dashed curves) were taken from Ref.\cite{JohnSatch96} .

In calculations of optical potentials,  the ``in-medium``  parameters $\sigma, \alpha, \beta$ of the $\pi N$ amplitude were varied by fitting  the corresponding $\pi A$  cross sections to experimental data at angles of scattering up to 80 degrees with the help of minimization of the $\chi^2$-function (15). These parameters are presented in Table 1, and their applications, when calculating cross sections, lead to about the same $\chi^2$ magnitudes for all the fitted folding OPs.

Table 2 shows the sets of 12  parameters of the Kisslinger-type potentials fitted  to the data of scattering at every energy.  When fitting these parameters, we started from those obtained in Ref.\cite{JohnSatch96} for  the pion scattering on the  ${^{16}}$O nucleus at a bombarding energy of 162 MeV and also on the ${^{208}}$Pb target at  162, 180 and 291 MeV. In our  case,  when fitting parameters to the data for a larger set of nuclei ${^{28}}$Si,${^{40}}$Ca,${^{58}}$Ni,${^{208}}$Pb in a wide range of energies  130 -291 MeV, we found out that for different target-nuclei but at the same energy of incident $\pi$ mesons one can use the same set of 12 parameters. With increasing incident energies, these  parameters  increase or decrease rather smoothly except the imaginary parts of the $c$ and $C$ values which change trends of their behaviour from increasing to decreasing at energies of about 226-230 MeV.

Thus, one can conclude that in spite of the so different constructions of employed  potentials (the folding one and the Kisslinger type) and sometimes different behavior of them as functions of $r$, both OPs yield cross sections in reasonably good agreement with experimental data at energies of the 3-3 resonance. Indeed, as an example, in  Fig.5 we show the behaviour of the real and imaginary parts of optical potentials and the corresponding cross sections at three energies 162, 226  and 291 MeV for the $\pi^{-}$ scattering on $^{28}$Si calculated by using these both models. One can see that at energies of 162, 291 MeV the differential cross sections for both OPs are in good coincidence at angles of scattering up to about 70 degrees, and for scattering at 226 MeV they are close to each other at angles in the limits up to 50 degrees. At the same time, at all energies the real parts of the Kisslinger-type potentials (blue dots) are repulsive in the inner regions of  nuclei, while the corresponding real parts of  folding potentials (red solid curves) are  negative at the pion energy of 162 MeV  and become positive  at 226 and 291 MeV. At a high energy of 291 MeV  the real parts and the imaginary parts for both potentials are in close agreement to one another. Moreover, the behavior of the imaginary parts of both OPs is almost coincides in their surface region, and this fact means that this region plays a decisive role in explanation of experimental data at the first quarter of angles of scattering.

The "in-medium" effect on the parameters $\sigma$, $\alpha$, $\beta$ of the folding potential is revealed when one compares them to those but for the pion scattering on free nucleons. With the aim to express this effect more markedly, we show in  Fig.6  both the "in-medium" and  "free" parameters averaged over the charged $\pi$-meson data \,$\sigma_{av}=(\sigma_{\pi^{+}}+\sigma_{\pi^{-}})/2$, $\alpha_{av}=(\alpha_{\pi^{+}}+\alpha_{\pi^{-}})/2$, \, $\beta_{av}=(\beta_{\pi^{+}}+\beta_{\pi^{-}})/2$ in their dependence on the $\pi$-meson energy in the region of the 33-resonance. Below, in Fig.6 we use the green and black circles and lines for the corresponding description of pion scattering on free nucleons, while the red squares and blue triangles and lines are for the pion scattering on the bounded nucleons in ${^{28}Si}$ and ${^{40}Ca}$ nuclei, respectively. In Fig.6a one can see the total $\pi N$ cross sections $\sigma_{av}$ obtained from analysis of the pion scattering data of $\pi+ {^{28}}$Si (red squares) and $\pi+ {^{40}}$Ca (green triangles). They repeat the general behavior of the  $\pi p$ cross section (green circles) in the  (33)-resonance energy region for the pion scattering on free nucleons \cite{Carter71} . However, in the case of  scattering on free nucleons, their cross sections (green curve) turn out to be about 1.6 times larger than those for pion scattering on  bounded nuclear nucleons. This effect of attenuation of the $\pi N$ interaction in nuclear matter might be due to the Pauli exclusion principle when the incident pion  interacts with the nucleon blocked in the nuclear shell.

In Fig.6b, one can see the energy dependence of the parameter $\alpha_{av}=ReF_{av}(0)/ImF_{av}(0)$, where $F_{av}(0)$ is the charge averaged  $\pi N$ amplitude of scattering (4) at zero angles. For the pion scattering on the  bound nucleons the $\alpha_{av}$ parameter remains to be positive at energies up to about 230 MeV, but for scattering on free nucleons (green circles)\cite{Arndt85} it becomes negative already at about 170 MeV. With the aim of understanding the physical meaning of this effect, one can consider the $\pi N$ potential as $V(r)=(v_0+iw_0)\exp(-r^2/2\beta)$ which yields the scattering amplitude   in the Born approximation $F_{\pi N}(q)= c(-v_0-iw_0) \exp(-\beta q^2/2)$, where $c=(m/{\hbar}^2)\,\sqrt{2\pi}\,{\beta}^{3/2}$. Thus, comparing this amplitude with the general expression in the form of (4) $F_{{\pi}N}(q)\,={k\over 4\pi}(\alpha\sigma\,+ i\sigma)\exp(-\beta q^2/2)$, one can see that the sign of the $\alpha$ parameter is opposite to that of the potential $v_0$. Therefore, for the attractive potential $v_0=-|v_0|$ of the $\pi N$ interactions one has the positive parameter $\alpha$, as is seen in Fig.6b in the region of the pion kinetic energies  up to about 230 MeV. At higher energies this parameter $\alpha$ is negative. This means that the negative potential of interaction of a pion  with the bound nucleon decreases with an increase in its kinetic energy $T$ and becomes repulsive at about 230 MeV.

In Fig.6c, one  finds that, for example, for the $\pi+ {^{28}}$Si scattering (red boxes) at energies of about 170 MeV near the 33-resonance maximum, the "slope" parameter $\beta_{av}$ in the amplitude (4) of pion scattering on the bound nuclear nucleon is about 1.6 times smaller than that for the pion scattering on a ''free'' nucleon \cite{Locher71,Bano}. If for the quantitative estimation one takes the $\pi N$ interaction potential as $V(r)=V_0\exp(-r^2/2\beta)$, then one obtains the Born scattering amplitude in the form of  $\sim\exp(-\beta q^2 /2)$ as in eq.(4), and thus the corresponding rms radius of interaction will be ${\Re}=\sqrt{<r^2>}=\sqrt{3\beta}$. Therefore, a decrease $\beta$ of about 1.6 times leads to about 1.6 times decrease in the corresponding cross section $\sim{\pi{\Re}^2}=3\pi\beta$ of pion interaction with a bound nucleon in about 1.6 times.

%
\begin{table}[!ht]
\centering
\caption{Parameters of the Kisslinger-type potentials}
\begin{tabular}{|c|cl|cl|cl|}
 \hline
$E_\pi$ (MeV)& Re\,$b_0$& Re\,$b_1$ (fm) & Re\,$c_0$   & Re\,$c_1$ (fm$^3$) & Re\,$C_0$& Re\,$C_1$, fm$^6$\\
\hline
130 &-0.0592  & -0.1291  & 0.7479    & 0.4171    & 0.4919    & 1.8353   \\
162 &-0.0770  & -0.1219  & 0.4935    & 0.2754    & 0.4325    & 2.2229   \\
180 &-0.0853  & -0.1245  & 0.3671    & 0.2050    & 0.4369    & 2.2458    \\
226 &-0.1008  & -0.1214  & 0.0849    & 0.0478    & 0.5657    & 1.6316    \\
230 &-0.1018  & -0.1212  & 0.0634    & 0.0359    & 0.5850    & 1.5332    \\
291 &-0.1088  & -0.1189  & -0.2026   & -0.1123   & 1.0406    & -0.8569  \\
\hline
      & Im\,$b_0$& Im\,$b_1$ (fm)& Im\,$c_0$ & Im\,$c_1$ (fm$^3$) & Im\,$C_0$ & Im\,$C_1$ (fm$^6$)\\
\hline
130 & 0.0310 &  0.0007  &   0.3025  &   0.1502  &   2.0026  &   3.8615 \\
162 & 0.0403 &  0.0046  &   0.5525  &   0.2753  &   2.1163  &   4.4689 \\
180 & 0.0451 &  0.0071  &   0.6394  &   0.3185  &   2.1380  &   4.5148  \\
226 & 0.0559 &  0.0129  &   0.6712  &   0.3324  &   2.0230  &   3.5965  \\
230 & 0.0567 &  0.0134  &   0.6614  &   0.3272  &   2.0021  &   3.4478  \\
291 & 0.0675 &  0.0209  &   0.2617  &   0.1216  &   1.4690  &  -0.1836 \\
\hline
\end{tabular}
\end{table}

\section{Summary}
In conclusion, one can sum up that both the microscopic folding and local Kisslinger-type optical potentials provide good agreement with experimental data of the pion-nucleus elastic scattering at intermediate energies in the region of the $\pi N$ 3-3 resonance between 130 and 230 MeV.

Below we summarize the main results of the work:

1. The folding pion-nucleus optical potential was presented in the form depending  obviously on three parameters $\sigma$, $\alpha$, $\beta$ of the amplitude of pion scattering  on a bound nuclear nucleon.

2. Every set of these 3 parameters was obtained by fitting the calculated $\pi A$ cross section at every given $\pi$-meson energy in the region of the $\pi N$ \,\, $\Delta_{33}$ resonance and by using the target-nuclei ${^{28}}$Si, ${^{40}}$Ca, ${^{58}}$Ni, ${^{208}}$Pb. In particular, magnitudes of the $\pi N$ cross-sections $\sigma$ are about 1.6 times lower than the corresponding cross-sections of the scattering on a free nucleon.

3. Also, 12 parameters of the Kisslinger-type optical potential were fitted  to get agreement of the calculated and experimental $\pi^{\pm} A$ cross sections at the fixed energy but for the whole set of target-nuclei. In the case of the folding OP, its 3 parameters $\sigma,\,\alpha,\,\beta$  were fitted for the every considered $\pi^{+} A$ and $\pi^{-} A$ cross sections to the experimental data of scattering on the selected target-nucleus $A$ at the given pion energy.

4. The  agreement of the calculated cross sections with the experimental data obtained for  both the folding and Kisslinger-type optical potentials takes place in spite of  that the Kisslinger OP has some radial variations in the inner part of a target-nucleus. At the same time, both potentials are close to each other in the surface region, and thus one can conclude that this region plays a decisive role in the scattering mechanism at the angles $\theta<80^\circ$.

5. Comparisons of the $\pi N$ cross section $\sigma^{free}$ for pion scattering on free nucleons  with the cross section $\sigma^{in}$ for scattering on a bound  nucleon in a nucleus in the 33-resonance energy region show that pion interaction with nucleons in nuclear matter is weaker than that with free nucleons. The behavior of the $\alpha$ parameter indicates that the refraction process of pion scattering on the bound nuclear nucleon  plays a decisive role at energies up to about $T^{lab} = 300$ MeV.

In conclusion, we note the important feature of the microscopic folding
pion-nucleus OP: it is constructed as the integral of a nucleus  density distribution function and an elementary amplitude of the $\pi N$ interaction. Thus, one has a possibility to study the effect of nuclear matter on the $\pi N$ amplitude, when the nucleon $N$ is not free one but bounded in a nucleus. On the other hand, comparisons of the folding  calculations for $\pi$A cross sections with  experimental data on elastic scattering data occur in agreement with those made by using the Kisslinger optical potential and, from this viewpoint, one can not reveal the advantages of one of them.  One can hope that this problem could be resolved when doing complex analysis of both elastic and inelastic data of scattering with excitations of the low-lying collective states of nuclei by using the folding and Kisslinger potentials together.

\section{Acknowledgments}
The work is partly supported by the Scientific Cooperation Program "ARE - JINR"


\begin{thebibliography}{99}
\bibitem{Faldt68} G.Faldt and T.E.O.Ericson, {Nucl.Phys} B8 (1968) 1.
\bibitem{Bjornenak70} K.Bjornenak, J.Finjord, P.Osland, A.Reitan, {Nucl.Phys} B20 (1970) 327;
 B22 (1970) 179.
\bibitem{Glauber59} R.J.Glauber, "Lectures in Theoretical Physics",(New York, Interscience, 1959)  Vol.1, p.315.
\bibitem{Czyz69} W.Czyz and L.C. Maximon, {Ann. Phys.(N.Y.)} 52 (1969) 59.
\bibitem{Charlton71} L.A.Charlton and J.M.Eisenberg, {Ann. Phys.} 63 (1971) 286.
\bibitem{Landau73} R.H.Landau, S.C.Phatak, F.Tabakin, {Ann. Phys.} 78 (1973) 299.
\bibitem{Kiss55} L.S.Kisslinger, {Phys. Rev.} 98 (1955) 761.
\bibitem{Miller74} G.A.Miller, {Nucl. Phys.} A 223 (1974) 477.
\bibitem{Liu81} L.C.Liu and S.M.Shakin, {Progress Part.Nucl.Phys.}  5 (1981) 207.
\bibitem{Khallaf2002} S.A.E.Khallaf and A.A.Ebrahim, {Phys. Rev.} C 62 (2000) 024603;
 C 65 (2002) 064605.
\bibitem{Satchler92} G.R.Satchler, {Nucl. Phys.} A 540 (1992) 533.
\bibitem{Kamalov87} M.Gmitro, S.S.Kamalov, R.Mach, {Phys. Rev.} 36 (1987) 1105.
\bibitem{KrellEr69} M.Krell and T.E.O.Ericson, {Nucl. Phys.} B 11 (1969) 521.
\bibitem{JohnSatch96} M.B.Johnson and G.R.Satchler, {Ann. Phys.} 248 (1996) 134.
\bibitem{LukZemLuk06} V.K.Lukyanov, E.V.Zemlyanaya, K.V.Lukyanov, {Phys. Atom. Nucl.} 69 (2006) 240.
\bibitem{LukZemLuk10} V.K.Lukyanov, E.V.Zemlyanaya, K.V.Lukyanov, K.M.Hanna, {Phys. Atom. Nucl.} 73 (2010) 1443.
\bibitem{LukZemLukZZ1214} V.K.Lukyanov, E.V.Zemlyanaya, K.V.Lukyanov, E.I.Zhabitskaya, M.V.Zhabitsky, {Phys. Atom. Nucl.} 77 (2014) 100; Preprint JINR P4-2012-105, JINR, Dubna, 2012.
\bibitem{Preedom79} B.M.Preedom, R.Corfu, J.P.Egger, P.Gretillat, C.Lunke, J.Piffaretti, E.Schwarz, J.Jansen, C.Perrin, {Nucl. Phys.} A 326 (1979) 385.
\bibitem{Olmer80} C.Olmer, D.F.Geesaman, B.Zeidman, S.Chakravarti, T.S.H.Lee, R.L.Boudrie, R.H.Siemssen, J.F.Amann, C.L.Morris, H.A.Thiessen, G.R.Burleson, M.J.Devereux, R.E.Segel, L.W.Swenson,  {Phys.Rev.} C 21 (1980) 254.
\bibitem{Geesaman81} D.F.Geesaman, C.Olmer, B.Zeidman, R.L.Boudrie, G.S.Blanpied, M.J.Devereux, G.R.Burleson, R.E.Segel, L.W.Swenson, H.A.Thiessen, {Phys.Rev.} C 23 (1981) 2635.
\bibitem{LukZemSlow04} V.K.Lukyanov, E.V.Zemlyanaya, B.S{\l}owinski, {Phys. Atom. Nucl.} 67 (2004) 1282; V.V.Burov, V.K.Lukyanov, Preprint P4-11098, JINR, Dubna, 1977.
\bibitem{LukZemLukMag17} V.K.Lukyanov, E.V.Zemlyanaya, K.V.Lukyanov, I.A.M.Abdul-Magead, {EPJ Web of Conferences}  138 (2017) 01019.
\bibitem{DeVries87} H. De Vries, C.W. De Jager,  C. De Vries, {Atomic Data and Nucl. Data Tables} 36 (1987) 495.
\bibitem{Gretillat81} P.Gretillat, J.P.Egger, J.F.Germond, C.Lunke, E.Schwarz, C.Perrin, B.M.Preedom, {Nucl. Phys.} A 364 (1981) 270.
\bibitem{DaoNP2000} Dao T.Khoa, G.R.Satchler, {Nucl. Phys.} A 668 (2000) 3.
\bibitem{LukZemLukMag20} K.V.Lukyanov, V.K.Lukyanov, E.V.Zemlyanaya and Ibrahim Abdulmagead, {Journal of Phys., Conf. Series} 1555 (2020) 012018.
\bibitem{PatPet2003} J.D.Patterson, R.J.Peterson, {Nucl. Phys.} A 717 (2003) 235.
\bibitem{Carter71} A.A.Carter, J.R.Williams, D.V.Bugg, P.J.Bussey, D.R.Dance, {Nucl. Phys.} B 26 (1971) 445.
\bibitem{Arndt85} R.A.Arndt, J.M.Ford, L.D.Roper, {Phys.Rev.} D 32 (1985) 1085.
\bibitem{Locher71} M.P.Locher, O.Steinmann, N.Straumann, {Nucl. Phys.} B 27 (1971) 598.
\bibitem{Bano} N.Bano and I.Ahmad, J. Phys. G 5 (1979) 39.
\end{thebibliography}
\end{document}